\documentclass[twocolumn,amsmath,amssymb,floatfix,pra,showpacs,footinbib,superscriptaddress]{revtex4}
\usepackage{amsmath,amssymb,natbib,bm,graphicx,psfrag}
\usepackage[ansinew]{inputenc}

\newcounter{fig}
\newenvironment{romenu}{\begin{list}{(\roman{fig})}{\usecounter{fig} \setlength{\labelwidth}{1.5cm}}}{\end{list}}

\newcommand{\MM}{\ensuremath{\langle M\rangle}}

\newcommand{\abs}[1]{\left|#1\right|}
\newcommand{\bra}[1]{\left\langle \, #1 \,\right|}
\newcommand{\ket}[1]{\left|\, #1 \, \right\rangle}

\newcommand{\be}{\begin{equation}}
\newcommand{\ee}{\end{equation}}

\DeclareMathOperator{\sgn}{sgn}

\DeclareMathOperator{\tr}{tr}

\begin{document}
\title{Proposal for generating and detecting multi-qubit GHZ states in circuit QED}
\author{Lev S.\ Bishop}
\affiliation{Departments of Physics and Applied Physics, Yale University, New Haven, Connecticut 06520, USA}
\author{L.\ Tornberg}
\affiliation{Department of Microtechnology and Nanoscience -- MC2, Chalmers University of Technology, SE-41296 Gothenburg, Sweden}
\author{D.\ Price}
\affiliation{Departments of Physics and Applied Physics, Yale University, New Haven, Connecticut 06520, USA}
\author{E.\ Ginossar}
\affiliation{Departments of Physics and Applied Physics, Yale University, New Haven, Connecticut 06520, USA}
\author{A.\ Nunnenkamp}
\affiliation{Departments of Physics and Applied Physics, Yale University, New Haven, Connecticut 06520, USA}
\author{A.\ A.\ Houck}
\affiliation{Department of Electrical Engineering, Princeton University, Princeton, New Jersey 08544, USA}
\author{J.\ M.\ Gambetta}
\affiliation{Institute for Quantum Computing and Department of Physics and Astronomy,
University of Waterloo, Waterloo, Ontario, Canada, N2L 3G1}
\author{Jens Koch}
\affiliation{Departments of Physics and Applied Physics, Yale University, New Haven, Connecticut 06520, USA}
\author{G.\ Johansson}
\affiliation{Department of Microtechnology and Nanoscience -- MC2, Chalmers University of Technology, SE-41296 Gothenburg, Sweden}
\author{S.\ M.\ Girvin}
\affiliation{Departments of Physics and Applied Physics, Yale University, New Haven, Connecticut 06520, USA}
\author{R.\ J.\ Schoelkopf}
\affiliation{Departments of Physics and Applied Physics, Yale University, New Haven, Connecticut 06520, USA}
\date{February 2, 2009}
\begin{abstract}
We propose methods for the preparation and entanglement detection of multi-qubit GHZ states in circuit quantum electrodynamics. Using quantum trajectory simulations appropriate for the situation of a weak continuous measurement, we show that the joint dispersive readout of several qubits can be utilized for the probabilistic production of high-fidelity GHZ states.
When employing a nonlinear filter on the recorded homodyne signal, the selected states are found to exhibit values of the Bell-Mermin operator exceeding 2 under realistic conditions. We discuss the potential of the dispersive readout to demonstrate a violation of the Mermin bound, and present a measurement scheme avoiding the necessity for full detector tomography.
\end{abstract}
\date{\today}
\maketitle

\section{Introduction\label{sec:intro}}
Performing measurements on separable input states is an elegant way to create entangled states of two or more qubits without the need for high-fidelity two-qubit gates \cite{duan_long-distance_2001,ruskov_entanglement_2003,sarovar_high-fidelity_2005,maunz_quantum_2007,blais_quantum-information_2007}. For this method to work, the measurement needs to be a \emph{joint readout} of several qubits, as single-qubit measurements are insufficient for entanglement generation. In circuit quantum electrodynamics (cQED) \cite{blais_cavity_2004,wallra_strong_2004}, the dispersive readout constitutes such a multi-qubit measurement.
The qubits, realized by superconducting charge qubits, are coherently coupled to the voltage inside the resonator and give rise to a state-dependent shift of  the resonator frequency.  Heterodyne or homodyne detection of the phase of microwave radiation transmitted through the resonator can thus be used to infer the state of the qubits.
Ideally, the corresponding measurement operator is simply a weighted sum of the qubits' Pauli operators $\sigma^z$ , where the weights are conveniently adjusted by the detunings of the respective qubits from the resonator frequency.

The idea of \emph{probabilistic state-preparation by measurement} has recently been applied to a 2-qubit cQED system by Hutchison et al.\ \cite{hutchison_quantum_2008}. Their theoretical study included the adverse effects of qubit relaxation and dephasing, and has shown the practical applicability of the method, even for realistic decay and decoherence rates as currently realized in cQED experiments. Here, we extend this method to the generation of multi-qubit Greenberger-Horne-Zeilinger (GHZ) states \cite{greenberger_going_1989}.
For superconducting qubit systems there have been successful demonstrations of Bell-state preparation \cite{steffen_measurement_2006,plantenberg_demonstration_2007,filipp_two-qubit_2008}, and proposals for creating GHZ states, mainly focusing on the generation via two-qubit gates and qubit-qubit interactions, see e.g.\ \cite{galiautdinov_maximally_2008,wei_generation_2006}. Instead of employing such entangling gates for generating a GHZ state, we propose a scheme tailored to cQED, consisting of one-qubit rotations and a dispersive measurement only.
Based on quantum trajectory simulations, we show that currently attainable values for qubit decoherence and decay allow for the creation of three-qubit GHZ states in cQED with high fidelity and high degree of entanglement. The degree of entanglement can be increased at the cost of lowering production rates.

In order to verify the production of the desired GHZ state, we propose to use a second dispersive readout.  Because the GHZ state is maximally entangled, this verification is related to proving the violation of a Bell-type inequality\cite{bell_problem_1966,aspect_experimental_1981}. However, proving such a violation in a \textit{loophole-free} fashion turns out to be a much more challenging task in cQED.
Given the required measurement time of hundreds of nanoseconds, space-like distances (in the sense of special relativity) are of the order of tens of meters and thus difficult to achieve in a cQED setup, and the dispersive readout is  in fact inherently nonlocal. Accepting that the locality loophole therefore cannot strictly be closed, we discuss the potential of the dispersive readout for observing quantum correlations in a 3-qubit GHZ state, as well as the potential for devising a factorizing measurement that is local in the no-signalling sense \cite{dieks_inequalities_2002}. Using quantum trajectory simulations including the measurement imperfections caused by qubit decay, we show that a convincing
violation of the Bell inequality would require a signal-to-noise ratio which is currently out of
experimental reach, but may be approached once efficient methods for protecting qubits from decay have
been devised, or with improvements in the noise performance of microwave amplifiers \cite{bergeal_analog_2008}.

Our paper is organized as follows: In the following section, we present the central idea of generating and detecting multi-qubit GHZ states by dispersive measurements, starting with the exposition of our proposal in the idealized situation of no qubit decoherence and decay. The rest of the paper is devoted to the consequences of measurement imperfections introduced by decay during the measurement process. In Section \ref{sec:model}, we specify the treatment of qubit decay and continuous homodyne detection using an effective stochastic master equation previously introduced in Ref.\ \cite{hutchison_quantum_2008}.
Quantitative results from solving this master equation for the situation of GHZ-state generation are presented in Section \ref{sec:generation}.
We describe different protocols for accepting or rejecting a generated state as a GHZ state, and show in particular that nonlinear filtering offers a significant advantage over simple boxcar filters.
In Section \ref{sec:detection} we discuss the detection of GHZ states within the dispersive measurement scheme and comment on the potential to violate a Bell-type inequality.
 Finally, our conclusions are presented in Section \ref{sec:conclusions}.

\section{Idealized preparation and detection of GHZ states\label{sec:idealized}}
In this section, we lay out the essential ideas behind the preparation and detection of GHZ states using the joint dispersive readout typical for cQED. To keep the discussion as clear as possible, our exposition in this section will ignore the adverse effect of qubit decay and decoherence, and other possible sources of measurement imperfections. We turn to the full discussion of the realistic situation including these effects in the following sections.

The GHZ state \cite{greenberger_going_1989} is the maximally entangled multi-qubit state of the form
\be\label{ghz}
\ket{\text{GHZ}}=\bigg( \bigotimes_{j=1}^N \ket{\downarrow}_j+\bigotimes_{j=1}^N \ket{\uparrow}_j\bigg)/\sqrt{2},
\ee
where $\ket{\cdot}_j$ denotes the state of qubit number $j$.
GHZ states have received much attention in the context of violation of Bell-type inequalities, see e.g. \cite{mermin_extreme_1990,mermin_quantum_1990,pan_experimental_2000-1,cabello_bells_2002,zhao_experimental_2003}, ruling out classical local-hidden-variable theories as a valid description of nature.

Partly, the beauty of the GHZ state lies in the fact that, in principle, violation of classicality can be proven with a single measurement of the corresponding Bell-Mermin operator $M$ \cite{mermin_extreme_1990}, see Eq.\ \eqref{merminop}. This has to be contrasted with the situation of the two-qubit CHSH scheme \cite{clauser_proposed_1969,clauser_experimental_1974}, where such a proof necessarily requires accumulating statistics. Key to this difference is the property of the GHZ state being an eigenstate not only of $M$, but also simultaneously of the measurable parity operators which sum up to $M$. The $N$-qubit GHZ state is known to violate a Bell-type inequality by an amount that grows exponentially in the number of qubits \cite{mermin_extreme_1990}.

\subsection{Preparation scheme}
In the dispersive readout of cQED, the measurement outcomes are inferred from the detection of the homodyne signal for the microwaves transmitted through the resonator. In the absence of qubit decay and decoherence, the probability distribution $p(s)$ for the integrated signal $s$ (see Sec.\ \ref{sec:model} for precise definition) takes the form of Gaussian peaks, which initially overlap strongly and separate with increasing measurement time $t$ \cite{blais_cavity_2004,gambetta_protocols_2007,majer_coupling_2007,filipp_two-qubit_2008,gambetta_quantum_2008}. In the limit of negligible overlap between peaks, the dispersive readout corresponds to a projective measurement of the operator  $A=\sum_j\delta_j\sigma^z_j$, where the weights $\delta_j=\chi_j/\bar\chi$ are the fractional contributions to the mean dispersive shift, $\bar\chi=\sum_j\chi_j/N$.
In detail, the preparation scheme can be described as follows:
\begin{figure}[b]
    \centering
    \psfrag{o}[c][][1.0]{$s\,[3t\Gamma_\text{ci}^{1/2}/4]$}
    \psfrag{x}[c][][1.0]{$s\,[t\Gamma_\text{ci}^{1/2}]$}
    \psfrag{y}[c][][1.0]{$p(s)$}
    \psfrag{n}[c][][0.8]{$\nu_1$}
    \psfrag{m}[c][][0.8]{$\nu_2$}
    \psfrag{a}[c][][0.7]{$\ket{\downarrow\downarrow\downarrow}$}
        \psfrag{e}[c][][0.7]{$\ket{\uparrow\uparrow\uparrow}$}
    \psfrag{b}[c][][0.7]{$
\begin{array}{c}
\ket{\downarrow\uparrow\uparrow}\\
\ket{\uparrow\downarrow\downarrow}
\end{array}$
}
\psfrag{d}[c][][0.7]{$
\begin{array}{c}
\ket{\uparrow\downarrow\downarrow}\\
\ket{\downarrow\uparrow\downarrow}
\end{array}$
}
\psfrag{c}[c][][0.7]{$
\begin{array}{c}
\ket{\downarrow\downarrow\uparrow}\\
\ket{\uparrow\uparrow\downarrow}
\end{array}$
}
    \psfrag{u}[c][][0.7]{$\ket{\downarrow\downarrow\downarrow}$}
    \psfrag{v}[c][][0.7]{$
\begin{array}{c}
\ket{\downarrow\downarrow\uparrow}\\
\ket{\downarrow\uparrow\downarrow}\\
\ket{\uparrow\downarrow\downarrow}
\end{array}$
}
    \psfrag{w}[c][][0.7]{$
\begin{array}{c}
\ket{\downarrow\uparrow\uparrow}\\
\ket{\uparrow\uparrow\downarrow}\\
\ket{\uparrow\downarrow\uparrow}
\end{array}$
}
    \psfrag{z}[c][][0.7]{$\ket{\uparrow\uparrow\uparrow}$}
        \includegraphics[width=0.9\columnwidth]{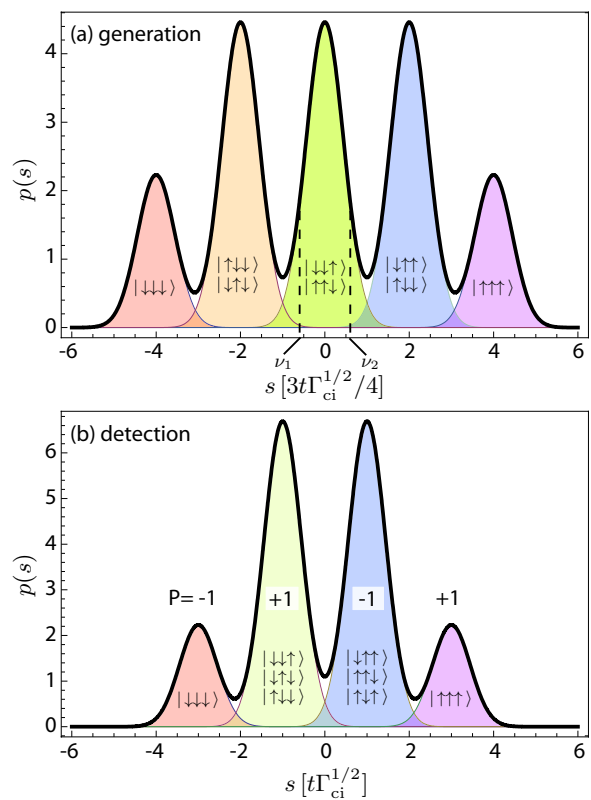}
    \caption{(Color online) Dispersive measurements employed for (a) generating a GHZ state, and (b) detecting the parity $\Pi=\prod_j\sigma^z_j$. Both panels show the probability density $p(s)$ for the integrated homodyne signal $s$ for the concrete example of a 3-qubit system. (a) For the generation of a 3-qubit pre-GHZ state, the dispersive shifts are fixed at ratios $\chi_1:\chi_2:\chi_3=1:1:2$. Ideally, the Gaussian peaks belonging to the 5 measurement results $\{\pm4,\pm2,0\}$ separate with increasing measurement time $t$ (here: $t=5/\Gamma_\text{ci}$), allowing for a reliable projective measurement when using appropriate thresholds, e.g.\ $\nu_1$, $\nu_2$ for the selection of the measurement outcome ``0". (b) The dispersive parity measurement  requires identical dispersive shifts, $\chi_1:\chi_2:\chi_3=1:1:1$. The four measurement  outcomes $a_i\in\{\pm3,\pm1\}$ then allow the inference of the parity value by $\Pi_i=-\sin(a_i\pi/2)$.\label{fig:figure1}}
\end{figure}
\begin{romenu}
\item Arrange all qubit detunings such that the system is dispersive, and mutual qubit detunings are large compared to the qubit-qubit interaction strengths. The initial state is the state with each qubit in its ground state, $\bigotimes_{j=1}^N\ket{\downarrow}_j$.
\item Perform $\pi/2$ rotations on each of the $N$ qubits, preparing the state $2^{-N/2}\bigotimes_{j=1}^N(\ket{\downarrow}_j+\ket{\uparrow}_j)$.
\item Keeping the system dispersive and mutual qubit detunings sufficiently large, adjust the qubit detunings such that their dispersive shifts assume the ratio,
\be
\chi_1:\chi_2:\ldots:\chi_{N-1}:\chi_N=1:1:\ldots:1:N-1,
\ee
 and perform a dispersive measurement. Ideally, this corresponds to a projective measurement of the observable $A=\sum_j\delta_j\sigma^z_j$. Conditioned on the measurement result being ``0", see Fig.\ \ref{fig:figure1}(a), we thus obtain the pre-GHZ state
\be
\ket{\text{pGHZ}} = \left( \ket{\downarrow\downarrow\ldots\downarrow\uparrow}+\ket{\uparrow\uparrow\ldots\uparrow\downarrow} \right)/\sqrt{2}.
\ee
\item In the final step, a $\pi$ rotation is applied to qubit $N$, yielding the GHZ state, Eq.\ \eqref{ghz}.
Alternatively, one may choose a different computational basis by interchanging the ``$\uparrow$" and ``$\downarrow$" labels for qubit $N$.
\end{romenu}
The necessary adjustment of the $\chi_j$ ratios is possible in cQED samples employing local flux-bias lines \cite{dicarlo}, which allow for the fine tuning of individual qubit frequencies.
We note that the scheme requires the resolution of only $\sim2N$ different peaks, which should be compared to
the need for application of $(N - 1)$ two-qubit gates for the preparation of the same GHZ state via gates, see e.g.
\cite{bodoky_production_2007}.

\subsection{Detection scheme}
Ideally, the confirmation of the GHZ state production and the verification of its quantum correlations proceed by a measurement of the Bell-Mermin operator \cite{mermin_extreme_1990},
\be\label{merminop}
M=2^{N-1}i\bigg( \prod_{j=1}^N \sigma^-_j- \prod_{j=1}^N \sigma^+_j \bigg).
\ee
For the $N$-qubit GHZ state, this operator takes on the value $2^{N-1}$, while local-hidden variable theories predict an outcome $\le 2^{N/2}$ if $N$ is even, and $\le 2^{(N-1)/2}$ if $N$ is odd \cite{mermin_extreme_1990} -- thus leading to a violation that grows exponentially in the qubit number.

In the general case, the Bell-Mermin operator is not amenable to a direct measurement. However, for $N$ qubits, it can be decomposed into $2^{N-1}$  $N$-qubit parity operators, which are more easily accessible by experiment, and the GHZ state is a simultaneous eigenstate of all the relevant parity operators. The specific form of the Bell-Mermin operator in the three-qubit case is given by
\be
M = \sigma^x_{1}\sigma^x_{2}\sigma^x_{3}-\sigma^x_{1}\sigma^y_{2}\sigma^y_{3}-\sigma^y_{1}\sigma^x_{2}\sigma^y_{3}-\sigma^y_{1}\sigma^y_{2}\sigma^x_{3}.
\ee
In the ideal case, one would perform the $2^{N-1}$ parity measurements, using a quantum non-demolition method on one and the same state and not requiring repeated measurements.

Since the dispersive readout does not realize exact parity measurements, we will accept the necessity to repeat measurements and acquire statistics. Instead of the parity, the dispersive readout can easily access the operator $A=\sum_j\sigma^z_j$, from which the value of the parity $\prod_j\sigma^z_j$ can be uniquely inferred, see Fig.\ \ref{fig:figure1}(b). Using single-qubit rotations mapping the appropriate $x$ and $y$ axes to $z$ \cite{kofman_analysis_2008}, all the required parities can be measured dispersively.

The crucial step thus consists in tuning all dispersive shifts to be identical. As before, this can be achieved by adjusting qubit detunings using local flux-bias lines. Compared to the setting employed for the GHZ state generation, it is in fact only the detuning of the $N$-th qubit that needs to be changed. Ideally, the measurement of $A$ then leads to the measurement outcomes $a_i\in\{\pm N,\pm(N-2),\ldots,\pm \ell\}$, terminating with $\ell=1$  if $N$ is odd and with $\ell=0$ if $N$ is even. The inferred parity outcomes simply alternate in sign according to $\Pi_i=-\sin(a_i\pi/2)$ for odd $N$ and by $\Pi_i=\cos(a_i\pi/2)$ for even $N$.
It is important to note that, while the number of required different measurements grows exponentially with $N$, the number of measurement outcomes that need to be resolved is given by $N+1$, only growing linearly with the qubit number.
This should be compared to the situation of a full state readout, which would require resolution of $2^N$ different peaks and dispersive shifts to be spread over an exponentially large frequency range, $\chi_j=2^j\chi_0$.

Both the generation and detection scheme will obviously suffer from qubit decoherence and decay. The subsequent sections take into account these effects and study quantitatively how the idealized proposal performs under more realistic conditions.

\section{Model\label{sec:model}}
For the generation and subsequent detection of a multi-qubit GHZ state we consider a cQED system \cite{blais_cavity_2004,wallra_strong_2004} consisting of three superconducting charge qubits  coupled to the fundamental mode of a microwave resonator.
The model of the system and notation follow those in Reference \onlinecite{hutchison_quantum_2008}.
Neglecting the possible influence of levels beyond the two-level approximation for the superconducting qubits, the system is described by a driven Tavis-Cummings Hamiltonian \cite{tavis_exact_1968}
\begin{align}\label{eq:hamiltonian1}
H =& \omega_r a^\dagger a  + \sum_j \frac{\omega_{q,j}}{2}\sigma^z_j + \sum_j g_j(a\sigma_j^+ + a^\dagger \sigma_j^- )\nonumber\\
&+ (a\epsilon^* e^{i\omega_m t} + a^\dagger \epsilon e^{-i\omega_m t}),
\end{align}
where we set $\hbar=1$, $\omega_r/2\pi$ denotes the resonator frequency, and $\epsilon$ the strength of the measurement drive. The qubit frequencies $\omega_{q,j}/2\pi$ are considered to be tunable individually, as realized by local flux-bias lines in recent cQED experiments \cite{dicarlo}. The qubit-resonator couplings are given by $g_j$, whose signs are determined by the location of the respective qubit  within the resonator. For concreteness, we will focus on the case of a $\lambda/2$ coplanar waveguide resonator, with two qubits placed close to one end, and the third qubit on the opposite end, leading to a relative sign $\sgn(g_1)=\sgn(g_2)=-\sgn(g_3).$

The system is to be operated in the dispersive regime, where $|\lambda_j| = |g_j|/|\omega_{q,j} - \omega_r|= |g_j/\Delta_j| \ll1$. Under these conditions the interaction term in Eq.\ (\ref{eq:hamiltonian1}) can be adiabatically eliminated \cite{blais_cavity_2004}, such that the effective Hamiltonian in the frame rotating with the measurement drive frequency $\omega_m$ reads
\be
H_{\mathrm{eff}} = \Delta_r a^\dagger a +  \sum_j  \frac{\omega_{q,j} - \chi_j}{2} \sigma_j^z + \sum_j \chi_j a^\dagger a \sigma_j^z + (\epsilon a + \epsilon^* a^\dagger),
\ee
where $\Delta_r = \omega_r - \omega_m$ is the detuning between measurement drive and resonator, and $\chi_j = g_j^2/\Delta_j$ denotes the dispersive shift due to qubit $j$ \footnote{We note that a structurally identical Hamiltonian is also obtained in the case of transmon qubits; merely the definition of the dispersive shifts $\chi_j$ is modified.}. Here, the qubit-qubit coupling $\sim J$ via virtual photons has been neglected, as is appropriate for sufficient detuning between qubits, $J\ll\abs{\Delta_j-\Delta_{j'}}$. The effects of qubit decay and cavity photon leakage are taken into account within a master equation description. Specifically, we include intrinsic qubit relaxation with rates $\gamma_{1j}$, Purcell-induced relaxation with rates $\gamma_{pj}$ \cite{purcell__1946,houck_controllingspontaneous_2008}, and photon decay from the cavity with rate $\kappa$. Pure dephasing can be strongly suppressed by proper design of the superconducting qubit \cite{schreier_suppressing_2008}, and will be neglected here. (We have checked that inclusion of pure dephasing at small rates, comparable to those achieved in \cite{schreier_suppressing_2008}, does not significantly alter our results.)

As demonstrated in Ref.\ \onlinecite{hutchison_quantum_2008}, one can dramatically simplify the resonator-qubit master equation and reach an effective master equation for the qubits only, given that photon decay is fast. Specifically, we assume that $\Delta_r=0$, and require
\be\textstyle
\kappa\gg \max\{\epsilon,\sum_j\abs{\chi_j}\}
\ee
Under these conditions, an analogous separation of qubit and resonator degrees of freedom can also be reached on the level of the stochastic master equation (SME), appropriate for the situation of continuous homodyne detection of the emitted microwave radiation \cite{hutchison_quantum_2008}. The effective SME for the qubit density matrix $\rho_J$ conditioned on the measurement record
\be
J(t) = \sqrt{\Gamma_\text{ci}}\sum_j\langle \delta_j\sigma^z_j\rangle + \xi(t)
\ee
is given by
\be\label{theSME}
\dot\rho_J=\mathcal{L}\rho_J + \sqrt{\Gamma_\mathrm{ci}}\xi(t)\mathcal{M}[\sum_j \delta_j \sigma_j^z] \rho_J,
\ee
where we are using notation identical to Ref.\ \onlinecite{hutchison_quantum_2008}: $\mathcal{M}[c]$ is the measurement operator given by
$
\mathcal{M}[c]\rho_J = (c-\langle c\rangle )\rho_J/2 + \rho_J (c-\langle c\rangle )/2
$,
$\xi(t)$ represents Gaussian white noise with zero mean and $\langle \xi(t)\xi(t')\rangle = \delta (t-t')$,  and $\Gamma_\text{ci}=\eta\Gamma_\text{m}$ denotes the effective measurement rate, reduced by an efficiency factor with respect to the maximum rate $\Gamma_\text{m}=64\bar{\chi}^2\abs{\epsilon}^2\kappa^{-3}$ \footnote{As in Ref.\ \cite{hutchison_quantum_2008}, we have checked that additional measurement-induced dephasing does not alter our results, and that the relevant parameter is the ratio of coherent information rate $\Gamma_\text{ci}$ and decay rates. As a result, we may set $\Gamma_d=\Gamma_\text{ci}/2$.}. The Liouvillian is defined as
\begin{align}
\mathcal{L}\rho=&-i\bigg[ \sum_j\frac{\omega_{qj}+\chi_j}{2}\sigma^z_j+\frac{4\bar\chi\abs{\epsilon}^2}{\kappa^2}\sum_j\delta_j\sigma^z_j,\rho\bigg]\\\nonumber
&+\sum_j(\gamma_{1j}+\gamma_{pj})\mathcal{D}[\sigma^-_j]\rho + \frac{\Gamma_\text{d}}{2}\mathcal{D}[\sum_j\delta_j\sigma^z_j]\rho
\end{align}
with the measurement-induced dephasing rate $\Gamma_\text{d}=\Gamma_\text{m}/2$,  and the usual dissipation superoperator $\mathcal{D}[A]\rho=A\rho A^\dag-\{A^\dag A,\rho\}/2$. Assuming mutually distinct qubit frequencies, we have treated the Purcell effect in the secular approximation. Specifically, we neglect the cross-terms in $\mathcal{D}[\sigma^-_1+\sigma^-_2-\sigma^-_3]$ which are
interference effects for radiation from different qubits, and which only become important if the qubit frequencies are sufficiently close, i.e.\ $\abs{\Delta_i-\Delta_j}\ll\gamma_{pi,j}$ \cite{majer_coupling_2007,hutchison_quantum_2008}.
In our case, Purcell-induced decay and intrinsic decay can be treated on the same footing, and in the following we will assume similar decay rates for all qubits and subsume them under the shorthand $\gamma=\gamma_{1j}+\gamma_{pj}$.
Finally, the integrated signal $s$ is simply given as the time integral of the measurement record for the full measurement time $t$, $s=\int_0^t dt'\,J(t')$.

\section{Preparation of the GHZ state under realistic conditions\label{sec:generation}}
We now turn to the situation of GHZ state preparation in the presence of qubit decay, which we study using quantum trajectory simulations based on the stochastic master equation \eqref{theSME}. Following the steps (i)--(iii) described in Section \ref{sec:idealized}, the system is initialized and dispersive shifts are adjusted for the measurement step. The interplay of measurement-induced dephasing, gradual state projection, and the simultaneous qubit decay are captured by the conditional density matrix $\rho_J$, where each simulation run generates a particular measurement record $J(t)$ up to a final measurement time $t$, corresponding to the experimentally accessible homodyne signal.

\begin{figure}
    \centering
        \includegraphics[width=1.00\columnwidth]{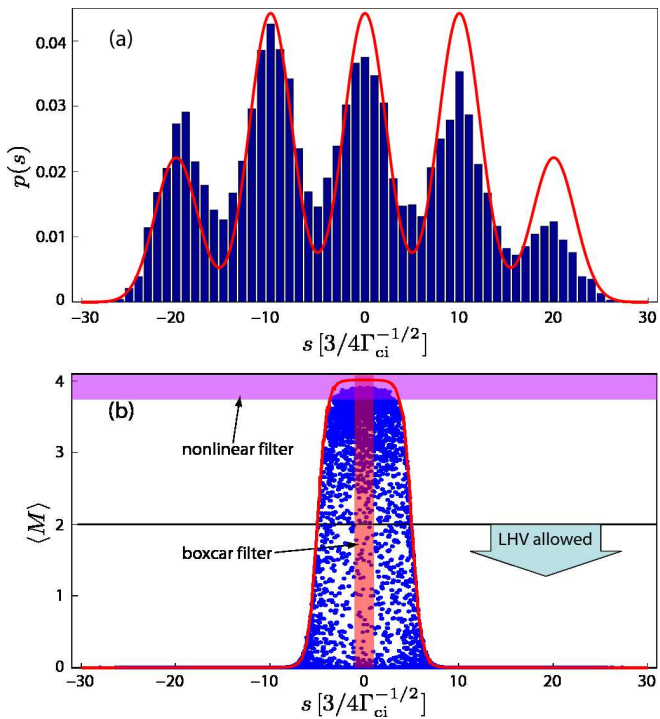}
    \caption{(Color online)  (a) Histogram of the integrated signal after a measurement time $t = 5/\Gamma_\text{ci}$, and  probability distribution $p(s)$ in the absence of any decay (red/gray curve). (b) Scatterplot (blue/gray dots) showing the correlation between the expectation value of the Mermin operator $\langle M \rangle$ and the integrated signal $s$ for $t = 5/\Gamma_\text{ci}$. Each point corresponds to one of 10\,000 trajectories. For comparison, the correlation in the ideal case of no decay is shown as the red/gray curve. The boxes indicate the action of the boxcar and the nonlinear filtering scheme. Parameters are chosen as $\Gamma_\text{d} = \Gamma_\text{ci}/2$,  $\gamma/\Gamma_\text{ci}= 1/35$ for $j=1,2,3$, and $\delta_1 = \delta_2 = 3/4$ and $\delta_3 = 3/2$.
\label{fig:figure2}}
\end{figure}

Since preparation of the correct pre-GHZ state is probabilistic (ideally, state generation succeeds with probability $P=1/4$ in the present case), one has to define a criterion (``filter") for success of preparation, and postselect the corresponding subensemble \cite{gambetta_protocols_2007}. In principle, the information available to the filter is the full measurement record. In the following, we will discuss two different filters, the linear boxcar filter and the  full nonlinear Bayesian filter and compare their performance in selecting high-fidelity GHZ states under realistic conditions.

The simple filter already outlined in Section \ref{sec:idealized} is the linear boxcar filter. It compresses each measurement record into a single number, the integrated signal $s=\int_0^t dt'\, J(t')$, and declares successful pre-GHZ state preparation whenever $s$ falls within the limits of appropriately chosen thresholds, $\nu_1\le s\le\nu_2$. Otherwise, the state is rejected.

The results for the integrated signal of many such measurements are conveniently plotted in form of a histogram, see Fig.\ \ref{fig:figure2}(a). When compared to the probability distribution expected in the ideal case of no decay, one observes that qubit decay leads to a distortion of the probability density with an overall shift of probability density towards the left-most peak, i.e., towards the signal associated with the ground state.
The shift is thus easily understood as a consequence of decay processes acting during the finite measurement time.

As a benchmark for the quality of the generated states and its correlation with the integrated signal, Fig.\ \ref{fig:figure2}(b) shows a scatterplot of the expectation value of the Bell-Mermin operator $\langle M \rangle$ versus the integrated signal for 10\,000 individual measurement trajectories. For comparison, the corresponding scatterplot in the ideal case of no decay is shown to collapse to a single curve. The scatter in the nonideal case results in trajectories of the same integrated signal, but very different values of $\langle M \rangle$, and thus in a significant number of falsely accepted states within the simple boxcar filtering.
\begin{figure}
    \centering
          \psfrag{t}[c][][1.0]{$t\,[\Gamma_\text{ci}^{-1}]$}
          \psfrag{J}[c][][1.0]{$J(t)\,[3\Gamma_\text{ci}^{1/2}/4]$}
        \includegraphics[width=1.00\columnwidth]{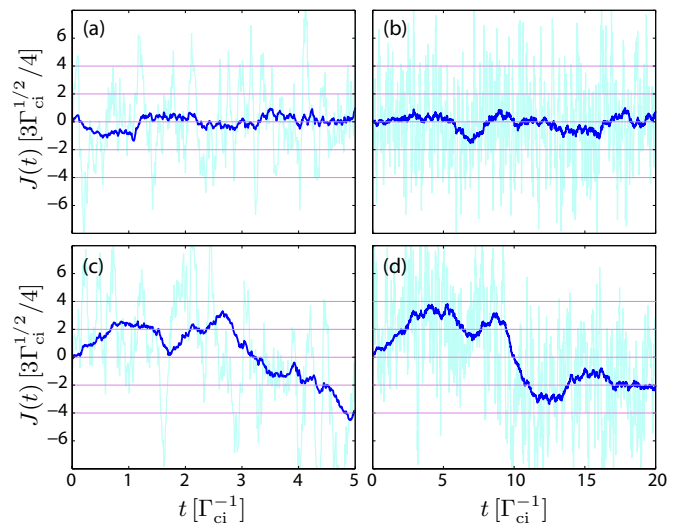}
    \caption{(Color online) Time traces of the signal $J(t)$ for individual quantum trajectories. The traces are smoothed over time $0.1 \Gamma_{\text{ci}}^{-1}$ (cyan/light gray) and $\Gamma_{\text{ci}}^{-1}$ (blue/dark gray). For (a) and (b) the expectation of the Mermin operator is large, $\langle M \rangle>3.9$, whereas for (c) and (d) it is small, $\langle M \rangle <0.1$. The horizontal lines indicate the values $J(t)$ would take on average for the integrated signal $s$ to be at the peaks of Fig.~\ref{fig:figure2}(a). All 4 traces are selected by boxcar filter on the integrated signal, such that they all lie close to the middle of the center peak. For (b),~(d) the relaxation is low, $\Gamma_{\text{ci}}/\gamma=142$,
    and trajectories with extremal values of $\langle M \rangle$ can be distinguished by eye.
For (a),~(c) the measurement time is
shorter and relaxation is faster $\Gamma_{\text{ci}}/\gamma=35$, nevertheless the nonlinear filter is still able to reliably estimate $\langle M \rangle$, as is demonstrated in Fig.~\ref{fig:fig4}.
\label{fig:fig3}}
\end{figure}

The essential mechanism for false acceptance of states is illustrated in Fig.\ \ref{fig:fig3}, showing the measurement record $J(t)$ as a function of time for four individual trajectories. Very roughly, the trajectories with integrated signal $s$ close to 0 can be divided into two categories: trajectories with measurement records $J(t)$ fluctuating around $J(t)=0$, see Fig.\ \ref{fig:fig3}(a),(b) and measurement records showing larger variations of $J(t)$ which accidentally average to $s=0$ upon integration. Trajectories of the first category correspond to the correct pre-GHZ state with high probability. On the other hand, an example from the second category consists of trajectories which, with high probability,  initially assume the state $\ket{\downarrow\uparrow\uparrow}$ with $\langle A \rangle =2$, and then suffer a decay process in qubit 3 at some intermediate time, thus transitioning to the state $\ket{\downarrow\uparrow\downarrow}$ with $\langle A \rangle =-2$, see Fig.\ \ref{fig:fig3}(c),(d).

This insight also points to a remedy for the boxcar filter. The full measurement record can, when spaced densely enough, be used to reconstruct the actual underlying quantum trajectory $\rho_J(t)$ in the following way: Given that the state before the onset of the measurement [see step (ii) in Sec.\ \ref{sec:idealized}] as well as the parameters entering the stochastic master equation are known with sufficient accuracy, one can successively determine the Wiener increments $dW(t) = \xi(t)dt$ from the measurement record. These, in turn, can then be used to propagate $\rho_J$ from the initial time to the measurement time $t$, and the resulting $\rho_J(t)$ encodes the expected value of the Bell-Mermin operator via $\langle M \rangle =\tr [\rho_J(t)M]$. This procedure corresponds to a nonlinear filter \cite{gambetta_protocols_2007}, with an acceptance criterion based on the value of $\langle M \rangle$ itself, see Fig.\ \ref{fig:figure2}(b).

\begin{figure}
    \centering
    \psfrag{MM}[][][0.8]{$\MM$}
    \psfrag{ap}[][][0.8]{acceptance probability}
    \psfrag{nla}[cl][cl][0.8]{$\Gamma_\text{ci}/\gamma=\infty$}
    \psfrag{nlb}[cl][cl][0.8]{$\Gamma_\text{ci}/\gamma=35$}
    \psfrag{nlc}[cl][cl][0.8]{$\Gamma_\text{ci}/\gamma=9$}
    \psfrag{nld}[cl][cl][0.8]{$\Gamma_\text{ci}/\gamma=3.5$}
    \psfrag{nle}[cl][cl][0.8]{$\Gamma_\text{ci}/\gamma=2$}
    \psfrag{snr}[][][0.7]{$\Gamma_\text{ci}/\gamma$}
    \psfrag{MN}[][][0.7]{$\MM$}
    \psfrag{bc}[Bl][Bl][0.5]{boxcar}
    \psfrag{nl}[Br][Br][0.5]{nonlinear}
    \includegraphics[width=0.95\columnwidth]{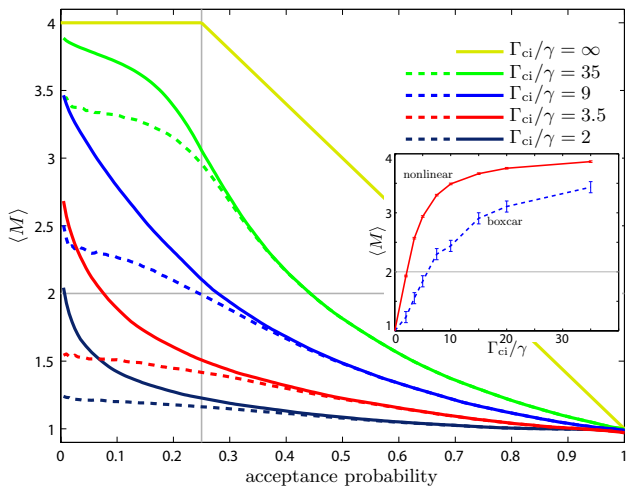}
    \caption{(Color online) Expectation value of the Mermin operator $ \langle M \rangle$ as a function of acceptance probability, for several ratios $\gamma/\Gamma_\text{ci}$; parameters are chosen as in Fig.\ 2. Solid (dashed) lines show the results using the nonlinear (boxcar) filter. (See text for details.) Using nonlinear filtering, the fraction of accepted trajectories with high $\langle M\rangle$-value can be substantially increased. For an acceptance probability $\lesssim 1/4$ the advantage of the nonlinear scheme becomes apparent. For each point, $\langle M \rangle$ is obtained by averaging over $20\,000$ trajectories and optimizing with respect to measurement time $t$ and boxcar thresholds. The inset shows the expectation value $\langle M \rangle$ as a function of the ratio $\Gamma_\text{ci}/\gamma$ for an acceptance probability of $1\%$.\label{fig:fig4}}
\end{figure}

The advantage of using the nonlinear filter is highlighted by Fig.\ \ref{fig:fig4}, which compares the performances of boxcar and nonlinear filter. For acceptance probabilities smaller than the ideally attainable $P=1/4$, we find that the nonlinear filter constitutes a significant improvement over the boxcar filter. Specifically, for ratios $\Gamma_\text{ci}/\gamma\alt4$ currently supported by experiments, the nonlinear filter will be crucial in order to reliably exceed the value $\langle M \rangle =2$, which is the relevant Mermin bound for violation of local-hidden variable theories in this case.
Figure \ref{fig:fig4} demonstrates that, when exploiting the trade-off between large expectation values of $\langle M \rangle$ and high acceptance probabilities, high-fidelity GHZ states can be prepared under realistic conditions.

\section{GHZ state detection under realistic conditions\label{sec:detection}}

\newcommand{\UU}{\ket{\Uparrow}}

\newcommand{\sg}[1]{\sigma^{x}_{#1}}

The measurement of the Bell-Mermin operator via parity detection, presented in Section \ref{sec:idealized}, requires the resolution of $\sim N$ peaks in the probability density $p(s)$ of the integrated signal.
While clearly advantageous relative to the resolution of $\sim 2^{N}$ peaks needed for a full readout, the parity detection remains difficult with current experimental parameters due to the qubit relaxation within the measurement time. In the following, we discuss a scheme that avoids this problem.

The key of this scheme lies in the fact that at low temperatures, decay into the state $\UU=\ket{\uparrow\uparrow\cdots\uparrow}$ is negligible. We note that this is similar to Kofman and Korotkov's use of the ``negative result outcomes"  to avoid the effects of measurement crosstalk in Bell tests using superconducting phase qubits \cite{kofman_analysis_2008}. False positive events in the detection of  the state $\UU$ can thus be suppressed by setting the acceptance threshold $\nu$ for the integrated homodyne signal sufficiently high.
Using this insight, we construct a measurement $B$ by assigning the measurement outcomes ``$0$", ``$1$" to the cases where the signal is respectively smaller or larger than a preset threshold.
We can describe $B$ in the language of generalized observables as a positive operator valued mapping (POVM) \cite{davies_quantum_1976}, by specifying its effects
\begin{align} \label{eq:e1}
E_1 &= \alpha \ket{\Uparrow}\bra{\Uparrow}, \\ E_0 &= \openone - E_1 .
\end{align}
Here, $\alpha=P_{\UU}(s>\nu)$  is the probability that the signal exceeds the threshold $\nu$ given the system was prepared in $\UU$. This probability is set by the decay of the $\UU$ state during the measurement time, and is analogous to the detector efficiency in quantum optics.  As a result, $1-\alpha$ can be described as a ``false negative'' probability that the measurement fails to detect a valid $\UU$ state. Experimentally, $\alpha$ can be determined by repeatedly preparing the system in $\UU$ (using single-qubit $\pi$ rotations), and subsequently performing the measurement. This procedure yields the expectation value $\langle \Uparrow|B|\Uparrow \rangle$, which is identical to the fraction of the cases where $s>\nu$, and hence to $\alpha$. In general, complete characterization of a POVM via detector tomography  \cite{lundeen_tomography_2009,luis_complete_1999,fiurek_maximum-likelihood_2001,dariano_quantum_2004} requires many measurements and a numerical optimization procedure to ensure the resulting POVM remains physical. Due to the simple structure of the measurement $B$, it may be conveniently characterized by determining only a single parameter $\alpha$.

 The measurement $B$ can now be combined with single-qubit rotations to determine the parity. We perform all combinations of $n$-qubit bit flips, $0\le n\le N$, and sum the measured $\langle B\rangle$ with relative sign $(-1)^n$. For clarity we specialize to the 3-qubit case, and define
 \begin{align} \label{eq:ff}
    f_{zzz}=&\langle B\rangle - \langle \sg1 B \sg1\rangle - \langle
\sg2 B \sg2\rangle -\langle \sg3 B \sg3\rangle \nonumber\\
     & + \langle \sg2\sg3 B \sg2\sg3\rangle + \langle \sg1\sg3 B \sg1\sg3\rangle + \langle \sg1\sg2 B\sg1\sg2\rangle\nonumber\\
     &- \langle \sg1\sg2\sg3 B \sg1\sg2\sg3\rangle .
\end{align}
The value of $f_{zzz}$ is proportional to the parity measured in the $z$-basis, $f_{zzz}=\alpha \langle \sigma_1^z\sigma_2^z\sigma_3^z\rangle$, with the proportionality constant being $\alpha$ as defined above.
It is straightforward to extend this scheme to the actual parities required for determining the value of the Bell-Mermin operator by prepending additional single-qubit rotations.

The expectation of the Bell-Mermin operator can now be related to the actual measurements via $F=\alpha\MM$, where
\begin{align}
    \label{eq:FF}
    F&=f_{xxx}-f_{xyy}-f_{yxy}-f_{yyx} .
\end{align}
Thus, the measurement of the 32 expectation values entering into $F$ and determination of $\alpha$ allow for the extraction of $\MM=F/\alpha$, with no restrictions on the qubits' decay rates \footnote{If decay is  fast, however, $\alpha$ may become so small that the time required for gathering sufficient statistics may become impractically long.}.

As explained in Section~\ref{sec:intro} the nature of the dispersive measurement prevents us in principle from a strict violation of a Bell-type inequality. However, in the limit where the measurement effects factorize into tensor products over the single-qubit Hilbert spaces, i.e.\ \begin{equation}
    E_{ijk}=E_{i}^{(1)}\otimes E_{j}^{(2)}\otimes E_{k}^{(3)} , \end{equation}
the measurement can be considered local in the sense of the no-signalling property \cite{dieks_inequalities_2002}. The effect defined in Eq.~\eqref{eq:e1} obeys such a factorization \begin{equation}
    E_1=\alpha \bigl(\ket{\uparrow}_1\bra{\uparrow}_1\bigr) \otimes
        \bigl(\ket{\uparrow}_2\bra{\uparrow}_2\bigr) \otimes
        \bigl(\ket{\uparrow}_3\bra{\uparrow}_3\bigr) .
\end{equation}
Similarly, the rotated measurements entering into $F$ factorize in this sense, provided the rotations themselves also factorize.
\begin{figure}
            \centering
                  \psfrag{a}[c][][1.1]{$1-\alpha$}
                  \psfrag{b}[c][][1.1]{$\beta$}
                  \psfrag{g1}[c][][1.0]{$\infty$}
                  \psfrag{g2}[c][][1.0]{$20$}
                  \psfrag{g3}[c][][1.0]{$\Gamma_\text{ci}/\gamma=5$}
      \psfrag{x}[c][][1.0]{threshold $\nu\,[\Gamma_\text{ci}^{1/2}/t]$}
      \psfrag{y}[c][][1.0]{$1-\alpha,\,\beta$}
      \includegraphics[width=0.9\columnwidth]{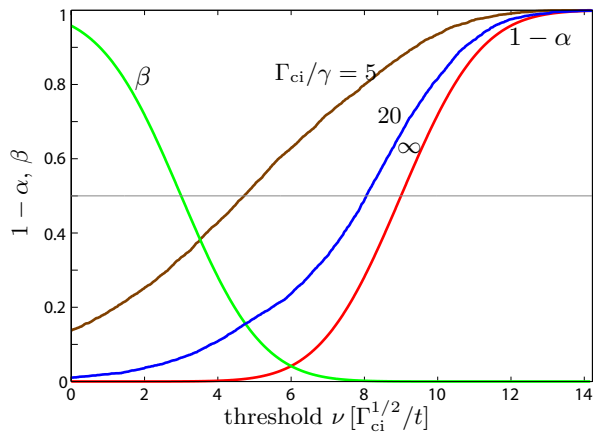}
      \caption{(Color online) False negative probability $1-\alpha$ and worst-case value for the false positive probability $\beta$ (definition see text) versus threshold $\nu$. The horizontal line indicates the necessary constraint on $\alpha$ to violate the Mermin inequality. The measurement time is chosen as $t = 3 / \Gamma_{\text{ci}}$.
 \label{fig:figure5}}
\end{figure}
This additional requirement holds not only for perfect single-qubit rotations \cite{kofman_analysis_2008}, but also for imperfect rotations, as long as there is no coupling or crosstalk between qubits during the rotation pulse. For example, independent single-qubit relaxation processes during a finite-duration rotation pulse do not spoil the factorization property. By contrast, a rotation of qubit $b$ caused by a rotation pulse on qubit $a$ no longer factorizes. In the following, we will assume that such crosstalk is negligible. In that case, the argument of Mermin applies, which states that a local hidden variable theory has bounds on the allowed $F$, $-2\le F\le 2$ \cite{mermin_extreme_1990}. Meanwhile quantum mechanics allows for $\MM=4$ and hence if $\alpha>1/2$ there is the possibility to violate Mermin's version of the Bell inequality.

Figure \ref{fig:figure5} shows the variation of the false negative probability $(1-\alpha)$ with threshold $\nu$,
so that the required threshold for $\alpha>1/2$ can be read off.
Since the derivation of the Bell inequality required factorization of measurement effects, we estimate the corrections to Eq.~\eqref{eq:e1}. In our case, the largest correction will be due to misidentification of states from the subspace $\{\ket{\downarrow\uparrow\uparrow},
\ket{\uparrow\uparrow\downarrow},\ket{\uparrow\downarrow\uparrow}\}$,
for which $A=\sum_j \sigma^z_j=1$. We put an upper bound on this misidentification probability $\beta$ by assuming that there is no decay out of this subspace and thus assume that $P_{A=1}(s)$, the distribution of the homodyne signal arising from this subspace, is Gaussian. Under these conditions, one obtains a worst-case estimate of the  ``false positive'' probability $\beta$ as a function of $\nu$.

Figure~\ref{fig:figure5} shows that with a low rate of qubit decay, $\Gamma_\text{ci}/\gamma=20$, we find $\alpha>1/2$ and a low probability of false positives, $\beta\simeq 0.002$, meaning that a meaningful violation of a Bell-type inequality should be possible. Conversely, for a more realistic rate of qubit decay $\Gamma_\text{ci}/\gamma=5$, the requirement $\alpha>1/2$ leads to significant false positive rates $\beta\simeq 0.16$, and factorization of $E_1$ breaks down. We note that the required $\Gamma_\text{ci}/\gamma\simeq 20$ for the violation of the Bell inequality is much more stringent than the experimenatally realistic $\Gamma_\text{ci}/\gamma\simeq 4$ that was shown in the previous section to be sufficient for producing states with $\MM>2$.

\section{Conclusions\label{sec:conclusions}}
In conclusion, we have presented a concrete proposal for efficient statistical production of multi-qubit GHZ states by dispersive measurement in a cQED setup, taking into account the realistic conditions of decoherence and decay. Our proposal is based on the possibility of adjusting the dispersive shifts of individual qubits, which effectively modifies the measurement operator and allows for the generation of entanglement starting from separable input states.
Our simulations show that even with experimentally achievable values of $2<\Gamma_\text{ci}/\gamma<4$ it is possible to achieve a $1\%$ efficiency in preparing states with values of the Bell-Mermin operator exceeding its classical bound, $\langle M \rangle>2$.

By using the global dispersive measurement in the same setup, we have also proposed a scheme for implementing parity measurements on the prepared state. Using these measurements, we have studied the sufficient conditions for verifying that such states indeed violate the Bell-Mermin inequality. We find that a ratio of $\Gamma_\text{ci}/\gamma=20$ (essentially identical to the signal-to-noise ratio) will be sufficient to observe a violation of the Mermin bound. While this ratio is larger than currently demonstrated, we are optimistic that the present limits on detector efficiencies in semiconductor amplifiers (1/20 of the quantum limit) may be improved by using superconducting pre-amplifiers \cite{bergeal_analog_2008}.
It would be interesting in the future to theoretically explore the possibility to use the full detector tomography for violation of the Bell-Mermin inequality. This could reveal the necessary conditions from the measurement setup for disproving local-hidden theories, putting less stringent constraints on experimental capabilities.

\begin{acknowledgments}
This work was supported in part by Yale University via a Quantum Information and Mesoscopic Physics Fellowship (JK),
by CIFAR, MITACS and ORDCF (JMG),
by LPS/NSA under ARO Contract No.\ W911NF-05-1-0365,
the NSF under Grants Nos.\ DMR-0653377 and DMR-0603369, the European Commission
through IST-015708 EuroSQIP integrated project, and by the Swedish Research Council.
\end{acknowledgments}

\end{document}